\documentclass[aps,twocolumn]{revtex4}

\usepackage{bm}

\usepackage{graphicx}

\usepackage{amssymb}

\begin{document}



\title{Odd-frequency Superconductivity on a Quasi-one-dimensional Triangular Lattice in the Hubbard Model}


\author{Keisuke Shigeta$^{1}$}
\author{Seiichiro Onari$^{1}$}
\author{Keiji Yada$^{2}$}
\author{Yukio Tanaka$^{1}$}
\affiliation{$$$^{1}$Department of Applied Physics, Nagoya University, Nagoya 464-8603, Japan\\}
\affiliation{$$$^{2}$Toyota Physical and Chemical Research Institute, Nagakute-cho 480-1192, Japan}

\begin{abstract}
In order to clarify whether the odd-frequency superconductivity can be
 realized or not, we study a quasi-one-dimensional
 triangular lattice in the Hubbard model using the random phase
 approximation (RPA) and the fluctuation exchange (FLEX)
 approximation. We find that odd-frequency spin-singlet $p$-wave pairing can be enhanced on a quasi-one-dimensional isosceles triangular lattice. 
\end{abstract}


\maketitle
\section{Introduction}
It is known that two electrons form Cooper pairs with gap function
$\Delta_{\sigma_1\sigma_2}(i\varepsilon_n,{\bm k})$ in superconductors. 
In general, the gap function depends on the Matsubara frequency $\varepsilon_n$, the combination of spins $\sigma_1$ and $\sigma_2$, and the momentum ${\bm k}$.
In accordance with the Fermi-Dirac statistics, the sign of the gap function is reversed by the exchange of two electrons: $\Delta_{\sigma_1\sigma_2}(i\varepsilon_n,{\bm k})=-\Delta_{\sigma_2\sigma_1}(-i\varepsilon_n,-{\bm k})$.
Based on the symmetrical properties combinations of three dependences, symmetries of the gap functions can be classified into four groups;
(i) even parity in frequency space (even-frequency), spin-singlet, and even parity in momentum space (even-parity) labeled as ESE pairing state,
(ii) even-frequency spin-triplet odd-parity (ETO) pairing state,
(iii) odd-frequency spin-singlet odd-parity (OSO) pairing state, and
(iv) odd-frequency spin-triplet even-parity (OTE) pairing state.
Almost all of the superconductors including high-$T_{\rm C}$ cuprates belong to ESE pairing state. 
ETO pairing state is realized in some special superconductors like Sr$_2$RuO$_4$ or UPt$_3$.
In contrast to these even-frequency pairings, odd-frequency pairings are not 
familiar.
\par
In 1974, the possibility of realizing the odd-frequency pairing state
was first proposed by Berezinskii in the context of $^3$He, where the odd-frequency spin-triplet hypothetical pairing was discussed\cite{Berezinskii}.
After that, Vojta and Dagotto pointed the possible realization of odd-frequency spin-triplet $s$-wave (OTE pairing state) on a triangular lattice in the Hubbard model\cite{Vojta}. 
Recent detailed calculation by Yada \cite{Yada} has supported this result. 
Balatsky and Abrahams proposed an odd-frequency spin-singlet $p$-wave pairing (OSO pairing state)\cite{Balatsky,Abrahams}.
Odd-frequency  pairing has been studied on the Kondo lattice model
\cite{Zachar,Coleman1,Coleman2}. 
There are some experimental reports\cite{Zheng,Kawasaki}, which are consistent with the realization of the odd-frequency spin-singlet $p$-wave superconducting state (OSO pairing state) in Ce compounds\cite{Zheng,Kawasaki,Fuseya}.
\par
It has been clarified recently 
that odd-frequency pairing correlation, 
$i.e.$, pair amplitude is generated 
in inhomogeneous system like superconducting junctions
\cite{Tanaka1,Tanaka2,Tanaka21,Tanaka3,Tanaka31,Eschrig} or vortex core \cite{Vortex,Vortex1}.
It has been pointed out that OTE pair amplitude 
is induced in ferromagnet / superconductor junctions\cite{Bergeret,Bergeret1,Bergeret2,Bergeret3,Bergeret4,Bergeret5,Bergeret6} 
and diffusive normal metal / spin-triplet odd-parity 
superconductor junctions\cite{Tanaka3,Tanaka31}. \par

Stimulated by these preexisting works, 
we explore the situation, where odd-frequency pairing is stabilized.
In this paper, we focus on the superconductivity on a quasi-one-dimensional triangular lattice.
In order to clarify the dominant pairing state there, we solve the linearized Eliashberg's equation in the Hubbard model using the random phase approximation (RPA) and the fluctuation exchange (FLEX) approximation\cite{Bulut,Bickers,Kondo}.
\par
We have found that odd-frequency spin-singlet $p$-wave becomes dominant pairing on a quasi-one-dimensional triangular lattice.
In particular, this pairing becomes prominent on an isosceles triangular
lattice due to geometrical frustration.
\par
\section{Model and Formulation}
\begin{figure}[htbp]
\begin{center}
\includegraphics[height=6cm]{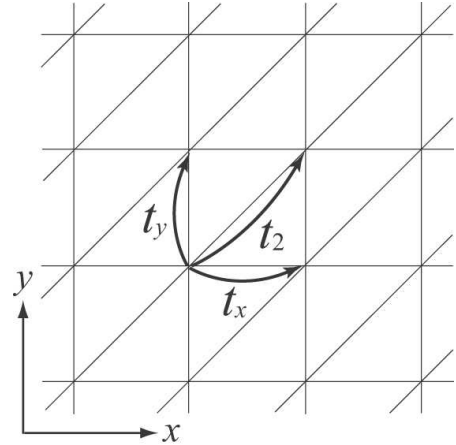}
\caption{A two-dimensional triangular lattice with transfer integrals $t_x$, $t_y$, and $t_2$.}
\label{figlattice}
\end{center}
\end{figure}

We start with a single-band Hubbard model on an anisotropic triangular lattice as shown in Fig. \ref{figlattice}, where $t_x$, $t_y$ and $t_2$ are transfer integrals along $x$, $y$ and diagonal directions, respectively.


The Hamiltonian is given by
\begin{eqnarray}
H&=&\sum_{\langle i,j\rangle,\sigma}\left(t_{ij}c_{i\sigma}^{\dag} c_{j\sigma}+h.c.\right)+\sum_{i}Un_{i\uparrow}n_{i\downarrow},
\end{eqnarray}
where $t_{ij}$ denotes the transfer integral between two sites $i$ and $j$. $\langle i,j\rangle$ is the combination of nearest and second-nearest neighbors.
$c_{i\sigma}^{\dag}$ ($c_{i\sigma}$) and $n_{i\sigma}=c_i^{\dag}c_i$ are creation (annihilation) and number operators, respectively. $U$ is the on-site Coulomb repulsion.
The band dispersion is given by
\begin{eqnarray}
\varepsilon_{\bm k}&=&-2t_x\cos k_x-2t_y\cos k_y-2t_2\cos\left(k_x+k_y\right).
\end{eqnarray}
In this study, the number of electrons per site is fixed to unity (half-filling).
\par
In this paper, we calculate Green's function $G(i\varepsilon_n,{\bm k})$ in two different ways; using (i) the RPA and (ii) the FLEX approximation.
\par
(i) In the RPA, the Green's function is given by $G(i\varepsilon_n,{\bm k})=(i\varepsilon_n-\varepsilon_{\bm k}+\mu)^{-1}$, where $\varepsilon_n$ is the Matsubara frequency of fermion given by $\varepsilon_n=(2n-1)\pi T$ with an integer $n$ and $\mu$ is the chemical potential. Using Green's function, the irreducible susceptibility is obtained as
\begin{eqnarray}
\chi_0(i\omega_m,{\bm q})=-\frac{T}{N}\sum_{n,{\bm k}}G(i\varepsilon_n,{\bm k})G(i\varepsilon_n+i\omega_m,{\bm k}+{\bm q}),\label{chi0}
\end{eqnarray}
where $\omega_m$ is the Matsubara frequency of boson given by $\omega_m=2m\pi T$ with an integer $m$ and $N$ is the number of sites.
The spin susceptibility is given by
\begin{eqnarray}
\chi_s(i\omega_m,{\bm q})&=&\frac{\chi_0(i\omega_m,{\bm q})}{1-U\chi_0(i\omega_m,{\bm q})},
\label{chis}
\end{eqnarray}
and the charge susceptibility is given by
\begin{eqnarray}
\chi_c(i\omega_m,{\bm q})&=&\frac{\chi_0(i\omega_m,{\bm q})}{1+U\chi_0(i\omega_m,{\bm q})}.
\label{chic}
\end{eqnarray}
\par
(ii) In the FLEX approximation, first the bare Green's function $G_0(i\varepsilon_n,{\bm k})=(i\varepsilon_n-\varepsilon_{\bm k}+\mu)^{-1}$ is calculated.
By substituting it into Green's function in Eq. (\ref{chi0}), we obtain the irreducible susceptibility.
The spin and charge susceptibilities are given by Eqs. (\ref{chis}) and (\ref{chic}), respectively.
Using the susceptibilities, the effective interaction is given by
\begin{widetext}
\begin{eqnarray}
V_n(i\omega_m,{\bm q})&=&\frac{3}{2}U^2\chi_s(i\omega_m,{\bm q})+\frac{1}{2}U^2\chi_c(i\omega_m,{\bm q})-U^2\chi_0(i\omega_m,{\bm q}).
\end{eqnarray}
\end{widetext}
Then, we calculate the self-energy
\begin{eqnarray}
\Sigma(i\varepsilon_n,{\bm k})&=&\frac{T}{N}\sum_{m,{\bm q}}V_n(i\omega_m,{\bm q})G(i\varepsilon_n-i\omega_m,{\bm k}-{\bm q}).
\end{eqnarray}
After using the Dyson equation
\begin{eqnarray}
G^{-1}(i\varepsilon_n,{\bm k})&=&G_0^{-1}(i\varepsilon_n,{\bm k})-\Sigma(i\varepsilon_n,{\bm k})
\end{eqnarray}
we obtain new Green's function.
The self-consistent iterations are repeated until the sufficient convergence is attained.
Thus, we obtain Green's function in which the self-energy is taken into account.
\par
Using Green's function, the spin susceptibility, and the charge
susceptibility obtained in the RPA or the FLEX approximation, we solve the linearized Eliashberg's equation.
The effective pairing interactions for spin-singlet and spin-triplet channel are given by
\begin{eqnarray}
V_a^s(i\omega_m,{\bm q})&=&U+\frac{3}{2}U^2\chi_s(i\omega_m,{\bm q})-\frac{1}{2}U^2\chi_c(i\omega_m,{\bm q}),
\\
V_a^t(i\omega_m,{\bm q})&=&-\frac{1}{2}U^2\chi_s(i\omega_m,{\bm q})-\frac{1}{2}U^2\chi_c(i\omega_m,{\bm q}),
\end{eqnarray}
respectively.
By substituting them into the linearized Eliashberg's equation for spin-singlet (spin-triplet) channel
\begin{widetext}
\begin{eqnarray}
\lambda\Delta(i\varepsilon_n,{\bm k})&=&-\frac{T}{N}\sum_{m,{\bm k}'}V_a^{s(t)}(i\varepsilon_n-i\varepsilon_m,{\bm k}-{\bm k}')G(i\varepsilon_m,{\bm k}')G(-i\varepsilon_m,-{\bm k}')\Delta(i\varepsilon_m,{\bm k}'),
\label{eliash}
\end{eqnarray}
\end{widetext}
the gap function $\Delta(i\varepsilon_n,{\bm k})$ and eigenvalue $\lambda$ are obtained.
Superconducting transition temperature $T_{\rm C}$ corresponds to the temperature where $\lambda$ reaches unity.
Thus, we consider that the larger the value of $\lambda$ becomes, the
more stable the superconductivity becomes.
In this paper, we take $N=N_x\times N_y=128\times64$ $\bm{k}$-point
meshes. The Matsubara frequencies $\varepsilon_n$ and $\omega_m$ have
values from
$-(2N_c-1)\pi T$ to $(2N_c-1)\pi T$ and from $-2N_c\pi T$ to $2N_c\pi T$, respectively, with $N_c=2048$.
\par
\section{Result}
First, we study superconducting state on a quasi-one-dimensional triangular lattice based on the RPA.
In our model, a quasi-one-dimensional triangular lattice is represented by choosing the values of $t_y$ and $t_2$ much smaller than that of $t_x$.
The quasi-one-dimensional direction is parallel to $x$-axis.
\par
\begin{figure}[htbp]
\begin{center}
\includegraphics[height=6cm]{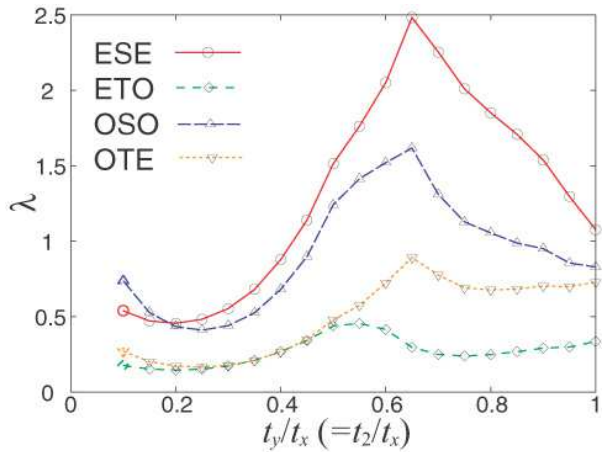}
\caption{(Color online) $t_y$ dependence of $\lambda$ for each pairing state, fixing $t_y=t_2$, obtained by the RPA with $S_t=0.97$ and $T/t_x=0.04$.}
\label{fig1}
\end{center}
\end{figure}

In order to clarify how symmetry of gap function depends on the dimension 
of the lattice, we gradually change the lattice structure varying $t_y$ and $t_2$ from a
two-dimensional regular triangular one ($t_x=t_y$) to a quasi-one-dimensional
triangular one ($t_x>t_y$).
Here, we choose $t_y=t_2$ for simplicity.
In this case, the lattice structure is an isosceles triangular one, in which spin frustration in the directions of $t_y$ and $t_2$ exists.
$\lambda$ of each pairing state changes as shown in Fig. \ref{fig1},
where Stoner factor $(S_t=U\chi_0^{Max})$ is fixed to $0.97$ by tuning the value of $U$.
Here, $\chi_0^{Max}$ denotes the maximum value of $\chi_0$.
We see that
OSO pairing state is dominant on a quasi-one-dimensional triangular lattice ($t_y/t_x\sim0.1$), while ESE pairing state is dominant in almost 
all of the region.
Hereafter, we focus on OSO pairing state.
\par
\begin{figure}[htbp]
\begin{center}
\includegraphics[height=6cm]{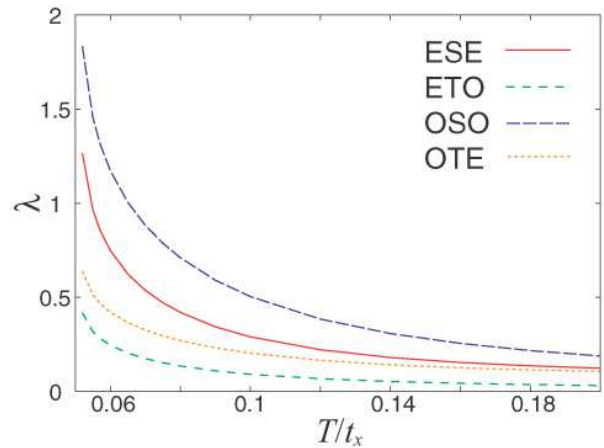}
\caption{(Color online) Temperature dependence of $\lambda$ for each pairing state by the RPA with $t_y/t_x=t_2/t_x=0.1$ and $U/t_x=1.6$.}
\label{fig3}
\end{center}
\end{figure}

In order to study details of the OSO pairing state on a quasi-one-dimensional triangular lattice, we calculate the
temperature dependence of $\lambda$ for $t_y/t_x=t_2/t_x=0.1$, and $U/t_x=1.6$.
As shown in Fig. \ref{fig3}, $\lambda$ of OSO pairing state increases
much above unity at low temperatures for $U/t_{x}=1.6$, 
where the Stoner factor $S_{t}$ becomes almost unity. 
The value of $S_t$ reaches unity at $T/t_x\sim0.05$.
This means that the OSO pairing state can be realized near the 
spin-density wave (SDW) phase. 
As in the case of ESE pairing, OSO pairing are 
are mediated by antiferromagnetic spin fluctuation\cite{Kondo} since the superconductivity appears near the SDW phase. 
%
\par
\begin{figure}[htbp]
\begin{center}
\includegraphics[width=8cm]{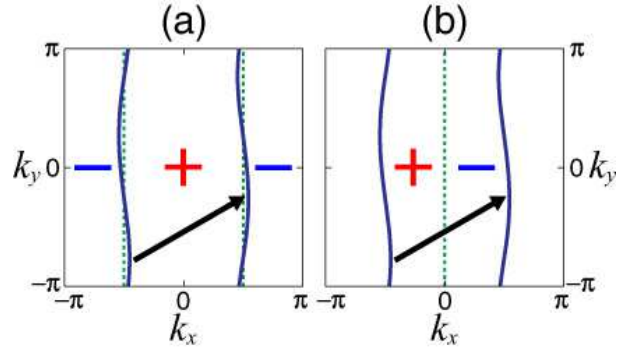}
\caption{(Color online) Momentum dependences of the gap functions for (a) ESE and (b) OSO
 pairing state by the RPA with $t_y/t_x=t_2/t_x=0.1$, $U/t_x=1.6$ and
 $T/t_x=0.06$, where the dashed and solid lines and the arrows represent the nodes of the gaps, the Fermi surfaces, and nesting vectors ${\bm Q}=(\pi,\pi/2)$,
 respectively.}
\label{fig4}
\end{center}
\end{figure}
\begin{figure}[htbp]
\begin{center}
\includegraphics[height=6cm]{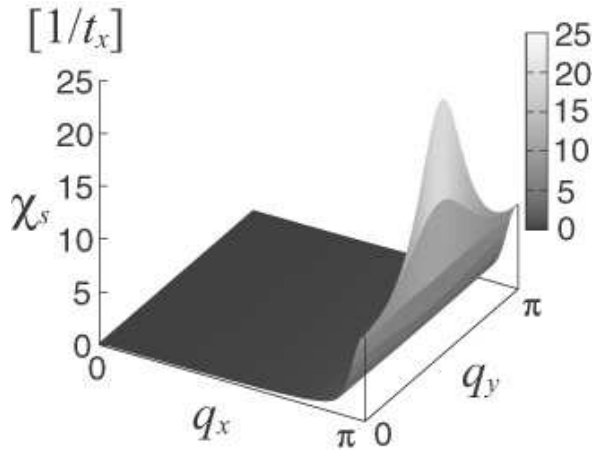}
\caption{Momentum dependence of the spin susceptibility by the RPA with $t_y/t_x=t_2/t_x=0.1$, $U/t_x=1.6$, and $T/t_x=0.06$.}
\label{fignest}
\end{center}
\end{figure}

The momentum dependences of the gap functions for the ESE and the OSO pairing states on a quasi-one-dimensional triangular lattice are
shown in Fig. \ref{fig4}(a) and (b), respectively, where the dashed and
solid lines and the arrows represent the nodes of the gaps, the Fermi surfaces,
and nesting vectors ${\bm Q}=(\pi,\pi/2)$, respectively.
We find that the momentum dependences of the ESE and the OSO gap functions can be approximated by $\cos k_x$ ($d$-wave) and $\sin k_x$ ($p$-wave), respectively.
Here, we denote $d$($p$)-wave since this gap function changes sign four (two) times on the Fermi surface.
It is noted that this $p$-wave has no nodes on the Fermi surface in the case of quasi-one-dimensional lattice ($t_y\lesssim 0.6$).
The shape of the Fermi surface becomes two lines with $k_x=\pm\pi/2$ in the absence of $t_y$ and $t_2$.
These two lines are bent into an S-shape by introducing $t_y$ and $t_2$.
It is noteworthy that these two "Fermi lines" are perfectly nested with a vector ${\bm Q}=(\pi,\pi/2)$ at half-filling with $t_y=t_2$.
Because of this, the spin susceptibility $\chi_s(i\omega_m,{\bm q})$ at ${\bm q}=(\pi,\pi/2)$ becomes strong as shown in Fig. \ref{fignest}.

It is known that on a two-dimensional
triangular
lattice, the antiferromagnetic
fluctuation works equally in the directions of $x$- and $y$-axes. Therefore, in real space, neighboring two electrons in these directions with antiparallel spins make a Cooper pair.
As a result, spin-singlet $d_{x^2-y^2}$-wave (ESE) becomes dominant.
On the other hand, the antiferromagnetic fluctuation along $x$-axis becomes dominant on a quasi-one-dimensional triangular lattice.
Then neighboring
two electrons along $x$-axis with antiparallel spins make a Cooper pair.
As a result, spin-singlet $d$-wave (ESE) and $p_x$-wave (OSO) make pairs in 
$x$-direction.

The gap function for ESE pairing state has nodes on the Fermi surfaces, while that for OSO pairing state is full-gap on the Fermi surfaces.
This point is relevant to following fact that OSO pairing state dominates over ESE pairing state for sufficiently small magnitude of $t_y$ and $t_2$.
\par

\begin{figure}[htbp]
\begin{center}
\includegraphics[height=6cm]{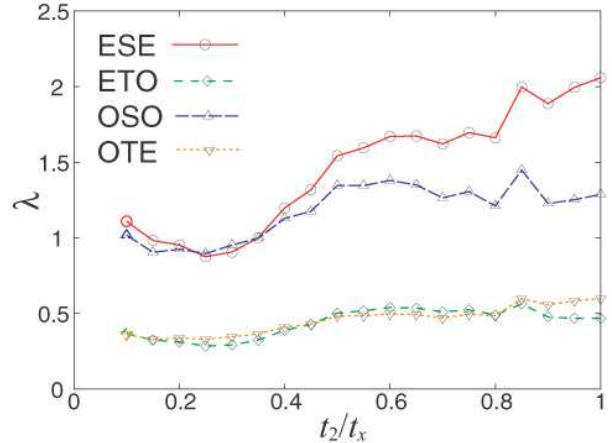}
\caption{(Color online) $t_2$ dependence of $\lambda$ for each pairing state by the RPA with $t_y/t_x=0.3$, $S_t=0.995$ and $T/t_x=0.05$.}
\label{fig2}
\end{center}
\end{figure}

We also explore the lattice with $t_y\ne t_2$ to clarify that $t_y=t_2$ is an indispensable condition for the realization of OSO pairing state.
We gradually change the value of $t_2$, fixing $t_y/t_x=0.3$.
The resulting $\lambda$ of each pairing state changes as shown in Fig. \ref{fig2}.
We see that OSO pairing state is relatively-enhanced
compared to ESE pairing state especially in the case of $t_y=t_2$.
This result is robust against changing value of $t_y$.
The condition with $t_y=t_2$ corresponds to  geometrical
frustration, which suppresses the antiferromagnetic fluctuation in these directions.
\par

\begin{figure}[htbp]
\begin{center}
\includegraphics[height=6cm]{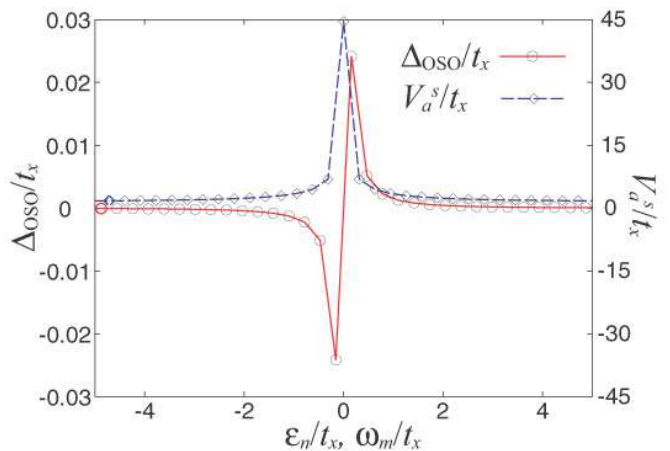}
\caption{(Color online) Matsubara frequency dependences of the gap function $\Delta_{\rm OSO}(i\varepsilon_n,{\bm k})$ at ${\bm k}=(\pi/2,0)$ and the effective pairing interaction $V_a^s(i\omega_m,{\bm Q})$ at ${\bm Q}=(\pi,\pi/2)$ for spin-singlet channel by the RPA with $t_y/t_x=t_2/t_x=0.1$, $S_t=0.95$, and $T/t_x=0.05$.}
\label{fig5}
\end{center}
\end{figure}

In the following, we study OSO pairing state focusing on frequency dependence. Matsubara frequency dependences of the gap function $\Delta_{\rm OSO}(i\varepsilon_n,{\bm k})$ for OSO pairing state at ${\bm k}=(\pi/2,0)$ and the effective pairing interaction $V_a^s(i\omega_m,{\bm Q})$ with ${\bm Q}=(\pi,\pi/2)$ for spin-singlet channel on a quasi-one-dimensional triangular lattice are shown in Fig. \ref{fig5}.
Near the SDW phase, $V_a^s(i\omega_m,{\bm Q})$ has a sharp peak at $\omega_m=0$ in frequency space.
After a simple transformation of the linearized Eliashberg's equation (\ref{eliash}), we obtain following relation,
\begin{widetext}
\begin{eqnarray}
\lambda&=&-\frac{T}{N}\frac{\sum_{n,m,{\bm k},{\bm k}'}V_a^{s(t)}(i\varepsilon_n-i\varepsilon_m,{\bm k}-{\bm k}')G(i\varepsilon_m,{\bm k}')G(-i\varepsilon_m,-{\bm k}')\Delta(i\varepsilon_m,{\bm k}')\Delta(i\varepsilon_n,{\bm k})}{\sum_{n,{\bm k}}|\Delta(i\varepsilon_n,{\bm k})|^2}.
\label{eliash2}
\end{eqnarray}
\end{widetext}
In Eq. (\ref{eliash2}), $\sum_{n,{\bm k}}|\Delta(i\varepsilon_n,{\bm k})|^2$, $G(i\varepsilon_m,{\bm k}')G(-i\varepsilon_m,-{\bm k}')$, and $V_a^{s}(i\varepsilon_n-i\varepsilon_m,{\bm k}-{\bm k}')$ are always positive.
Then negative (positive) sign of $\Delta_{\rm OSO}(i\varepsilon_m,{\bm k}')\Delta_{\rm OSO}(i\varepsilon_n,{\bm k})$ 
makes positive (negative) contribution to $\lambda$.
Due to the sharp peak of $V_a^s(i\omega_m,{\bm Q})$ at $\omega_m=0$, pair scattering from $\varepsilon_m$ to $\varepsilon_n$ with $\varepsilon_m= \varepsilon_n$ makes main contribution to $\lambda$.
The gap function $\Delta_{\rm OSO}(i\varepsilon_n,{\bm k})$ changes sign in the process of scattering from ${\bm k}'$ to ${\bm k}$ through the nesting vector ${\bm Q}=(\pi,\pi/2)$, which makes the main contribution to $\lambda$, 
in the momentum summation of the numerator.
However, scattering process for $\omega_m\ne 0$ suppresses the value of $\lambda$ since gap functions with positive and negative sign of 
$\varepsilon_n$  have opposite signs each other in OSO pairing.
\par
\begin{figure}[htbp]
\begin{center}
\includegraphics[height=6cm]{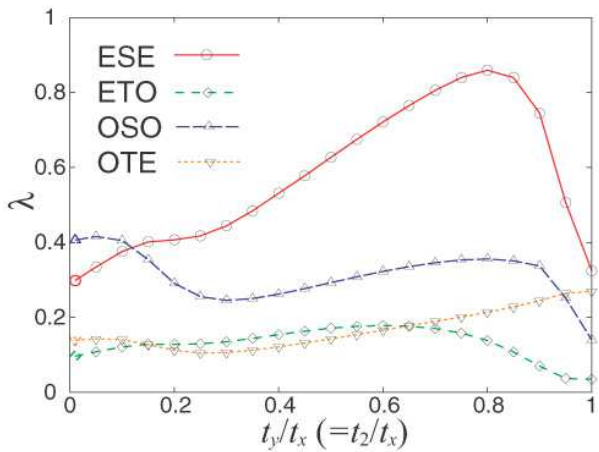}
\caption{(Color online) $t_y$ dependence of $\lambda$ for each pairing state fixing $t_y=t_2$ by the FLEX approximation with $S_t=0.97$ and $T/t_x=0.02$.}
\label{fig6}
\end{center}
\end{figure}

Next we calculate the value of $\lambda$ with the FLEX approximation in order to reveal how the above results is changed by the self-energy.
As in the case of the RPA, to clarify how the superconducting state depends on the dimensionality of the lattice, 
we gradually change the lattice structure from a two-dimensional regular triangular one ($t_x=t_y$) into a quasi-one-dimensional triangular one ($t_x>t_y$), fixing $t_y=t_2$.
$\lambda$ of each pairing state changes as shown in Fig. \ref{fig6}.
Compared to the result of the RPA in Fig. \ref{fig1}, the values of $\lambda$ are reduced.
However, similar to the case of the RPA for
$t_y/t_x\lesssim0.1$, the OSO pairing state still remains dominant.
\par
\begin{figure}[htbp]
\begin{center}
\includegraphics[height=6cm]{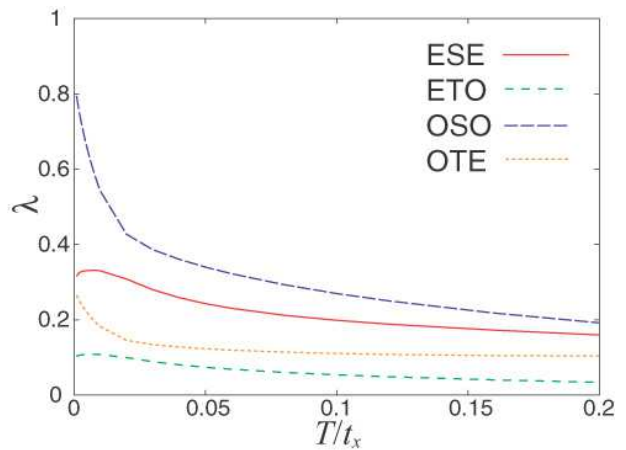}
\caption{(Color online) Temperature dependence of $\lambda$ for each pairing state by the FLEX approximation with $t_y/t_x=t_2/t_x=0.01$ and $U/t_x=2.5$.}
\label{fig7}
\end{center}
\end{figure}

We calculate the temperature dependence of $\lambda$ for $t_y/t_x=t_2/t_x=0.01$.
As shown in Fig. \ref{fig7}, the value of $\lambda$ for OSO pairing reaches 
up to $0.8$.
The present value of $\lambda$ is considerably high as compared to the
corresponding values studied by the FLEX approximation in other strongly
correlated systems\cite{NaCo-kuroki,NaCo-motizuki,SRO-onari,organic-kuroki,organic-kontani}.
Up to now, only the values of $\lambda$ obtained for high-$T_{\rm C}$ cuprates and exotic systems with disconnected Fermi surface exceed over the present value\cite{High-Tc,High-Tc1,High-Tc2,High-Tc3,disconnect-kuroki,disconnect-kuroki1,disconnect-onari}.
Momentum and frequency dependences of gap function are qualitatively similar to the one obtained by the RPA.
As regard the frequency dependence, there is only quantitative difference.
\par

\begin{figure}[htbp]
\begin{center}
\includegraphics[height=6cm]{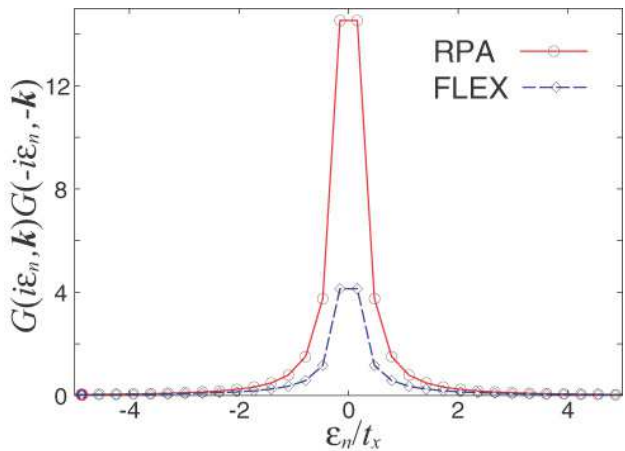}
\caption{(Color online) Matsubara frequency dependences of $G(i\varepsilon_n,{\bm k})G(-i\varepsilon_n,-{\bm k})$ at ${\bm k}=(\pi/2,0)$ by the RPA and the FLEX approximation with $t_y/t_x=t_2/t_x=0.1$, $S_t=0.95$, and $T/t_x=0.05$.}
\label{fig8}
\end{center}
\end{figure}

Matsubara frequency dependences of value of $G(i\varepsilon_n,{\bm k})G(-i\varepsilon_n,-{\bm k})$ at ${\bm k}=(\pi/2,0)$ obtained by the RPA and the FLEX approximation, which directly affect the value of $\lambda$ as noted in Eq. (\ref{eliash2}), are shown in Fig. \ref{fig8}.
$G(i\varepsilon_n,{\bm k})G(-i\varepsilon_n,-{\bm k})$ for the FLEX
approximation is smaller than that for the RPA since imaginary part of self-energy corresponds to damping of quasi-particles.
Thus, the presence of self-energy decreases the value of $\lambda$ in the FLEX approximation.
\par
\begin{figure}[htbp]
\begin{center}
\includegraphics[height=6cm]{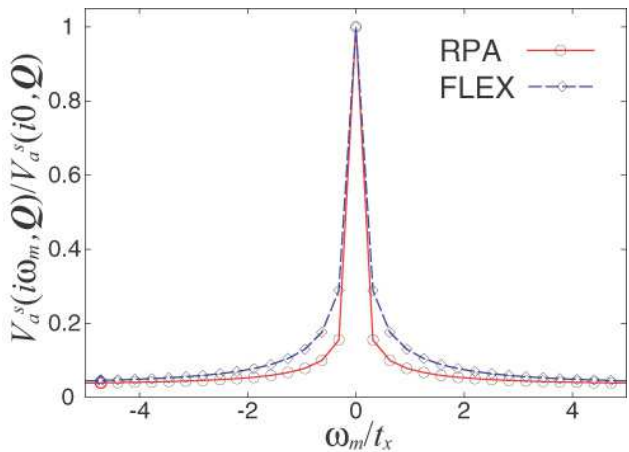}
\caption{(Color online) Matsubara frequency dependences of the normalized effective pairing interactions $V_a^s(i\omega_m,{\bm Q})/V_a^s(i0,{\bm Q})$ at ${\bm Q}=(\pi,\pi/2)$ for spin-singlet channel by the RPA and the FLEX approximation with $t_y/t_x=t_2/t_x=0.1$, $S_t=0.95$, and $T/t_x=0.05$.}
\label{fig9}
\end{center}
\end{figure}

Matsubara frequency dependences of the normalized effective pairing interactions $V_a^s(i\omega_m,{\bm Q})/V_a^s(i0,{\bm Q})$ at ${\bm Q}=(\pi,\pi/2)$ for spin-singlet channel in the RPA and the FLEX approximation are shown in Fig. \ref{fig9}.
A peak width of $V_a^s(i\omega_m,{\bm Q})$ for the FLEX approximation is broader than
that for the RPA in Matsubara frequency space due to the presence of the 
self-energy.
This broadness of the peak width of the effective pairing interaction
$V_a^s(i\omega_m,{\bm Q})$ relatively enhances the value of $V_a^s(i\omega_m,{\bm Q})$ with $\omega_m\ne 0$.
Subsequently, the scattering processes from positive $\varepsilon_n$ to negative $\varepsilon_n$ relatively increase the summation of Matsubara frequency 
in the numerator of Eq. (\ref{eliash2}).
Thus, the OSO pairing state is suppressed.
On the other hand, for the ESE pairing state, the above scattering processes do not suppress the value of $\lambda$ since the gap function has always same sign in frequency space.
As a result, the OSO pairing  is suppressed more significantly by the
self-energy as compared to the ESE pairing.
\par
\section{Conclusion}
We have studied symmetry of gap functions on a quasi-one-dimensional triangular lattice in the Hubbard model by solving the linearized Eliashberg's equation 
based on the RPA and the FLEX approximation. 
Surprisingly, odd-frequency spin-singlet $p_x$-wave 
(OSO pairing state), which is not familiar, is enhanced near 
the SDW phase. 
The OSO pairing state becomes prominent on an isosceles triangular lattice ($t_y=t_2$), where the geometrical frustration is the most significant.
Even if the self-energy is introduced by the FLEX approximation, 
above conclusions are not changed.
\par
The OSO pairing state is induced in the following way.
In real space, neighboring two electrons with antiparallel 
spins make a Cooper pair mediated by the antiferromagnetic spin 
fluctuation near the SDW phase.
The value of the effective pairing interaction $V_a^s(i\omega_m,{\bm Q})$ for spin-singlet channel at the nesting vector ${\bm Q}=(\pi,\pi/2)$ has a sharp positive peak at $\omega_m=0$ in the frequency space.
In this case, pair scattering with conserving 
frequency makes major contribution to the value of $\lambda$.
Therefore, sign inversion of gap function through the nesting vector ${\bm Q}=(\pi,\pi/2)$ in the momentum space enhances
the value of $\lambda$ in Eq. (\ref{eliash2}) 
because the effective pairing interaction for spin-singlet channel 
has a positive value.
This favors $p_x$-wave, which is full-gap on the Fermi surface.
In accordance with Fermi-Dirac statistics, spin-singlet $p_x$-wave can be interpreted as odd-frequency pairing.
These results in this paper suggest possibility of odd-frequency superconductivity realizing in bulk system.
\par
In the present paper, only on-site Coulomb interaction is considered. 
In the presence of off-site Coulomb interaction, it is known that 
SDW and CDW can compete each 
other in quasi-one-dimensional superconductor. 
In that case, 
competition between even-frequency spin-triplet $f$-wave pair and 
even-frequency spin-singlet $d$-wave pair has  been pointed by 
several theories\cite{Fuseya1,Fuseya11,Fuseya12,Fuseya13,Fuseya14}. 
It is a challenging issue to consider the possible realization of 
odd-frequency pairing in the presence of off-site Coulomb interaction. \par
Beside this problem, to clarify the superconducting properties 
of odd-frequency superconductor is an important problem. 
Since phase sensitive probes, $e.g.$, 
tunneling and Josephson effects, are crucial to identify the 
pairing symmetry in unconventional superconductors \cite{TK,TK1,TK2}, 
similar studies on odd-frequency superconductors 
become important \cite{Tanaka1,Linder,Linder1,Linder2}. 
It is necessary to calculate temperature dependence of 
energy gap function to reveal the superconducting properties. 

\par


\begin{thebibliography}{99}



\bibitem{Berezinskii}
V. L. Berezinskii,
JETP Lett. {\bf 20}, 287 (1974).

\bibitem{Vojta}
M. Vojta and E. Dagotto,
Phys. Rev. B {\bf 59}, R713 (1999). 

\bibitem{Yada}
K. Yada, S. Onari, Y. Tanaka, and K. Miyake, arXiv:0806.4241.

\bibitem{Balatsky}
A. Balatsky and E. Abrahams,
Phys. Rev. B {\bf 45}, 13125 (1992).

\bibitem{Abrahams}
E. Abrahams, A. Balatsky, D. J. Scalapino, and J. R. Schrieffer,
Phys. Rev. B {\bf 52}, 1271 (1995).

\bibitem{Zachar}
O. Zachar, S. A. Kivelson, and V. J. Emery,
Phys. Rev. Lett. {\bf 77}, 1342 (1996).

\bibitem{Coleman1}
P. Coleman, A. Georges, and A. M. Tsvelik,
J. Phys.: Condens. Matter {\bf 9}, 345 (1997).

\bibitem{Coleman2}
P. Coleman, E. Miranda, and A. Tsvelik,
Phys. Rev. B {\bf 49}, 8955 (1994).

\bibitem{Zheng}
G. Q. Zheng, N. Yamaguchi, H. Kan, Y. Kitaoka, J. L. Sarrao, P. G. Pagliuso, N. O. Moreno, and J. D. Thompson,
Phys. Rev. B {\bf 70}, 014511 (2004).

\bibitem{Kawasaki}
S. Kawasaki, T. Mito, Y. Kawasaki, G.-q. Zheng, Y. Kitaoka, D. Aoki, Y. Haga, and Y. Onuki, 
Phys. Rev. Lett. {\bf 91}, 137001 (2003).

\bibitem{Fuseya}
Y. Fuseya, H. Kohno, and K. Miyake,
J. Phys. Soc. Jpn. {\bf 72}, pp.2914 (2003).

\bibitem{Tanaka1}
Y. Tanaka, A. A. Golubov, S. Kashiwaya, and M. Ueda,
Phys. Rev. Lett. {\bf 99}, 037005 (2007). 

\bibitem{Tanaka2}
Y. Tanaka, Y. Tanuma, and A. A. Golubov,
Phys. Rev. B {\bf 76}, 054522 (2007).

\bibitem{Tanaka21}
Y. Tanaka, Y. Asano, and A. A. Golubov, 
Phys. Rev. B {\bf 77}, 220504(R) (2008). 


\bibitem{Tanaka3}
Y. Tanaka and A. A. Golubov,
Phys. Rev. Lett. {\bf 98}, 037003 (2007).

\bibitem{Tanaka31}
Y. Asano, Y. Tanaka, A. A. Golubov, and S. Kashiwaya,  
Phys. Rev. Lett. {\bf 99}, 067005 (2007)

\bibitem{Eschrig}
M. Eschrig, Lofwander, T. Champel, J. C. Cuevas, and G. Sch\"{o}n, 
J. Low Temp. Phys. {\bf 147}, 457 (2007).

\bibitem{Vortex}
T. Yokoyama, Y. Tanaka, and A.A. Golubov, 
Phys. Rev. B {\bf 78}, 012508 (2008).

\bibitem{Vortex1}
Y. Tanuma, N. Hayashi, Y. Tanaka, and A.A. Golubov,
arXiv:0808.0680.

\bibitem{Bergeret}
F. S. Bergeret, A. F. Volkov, and K. B. Efetov, 
Phys. Rev. Lett. \textbf{86}, 4096 (2001).

\bibitem{Bergeret1}
F. S. Bergeret, A. F. Volkov, and K. B. Efetov,
Rev. Mod. Phys. {\bf 77}, 1321 (2005).

\bibitem{Bergeret2}
M. Eschrig, J. Kopu, J. C. Cuevas, and G. Sch\"{o}n,
Phys. Rev. Lett. \textbf{90}, 137003 (2003).

\bibitem{Bergeret3}
T. Yokoyama, Y. Tanaka, and A. A. Golubov, 
Phys. Rev. B \textbf{75}, 094514 (2007).

\bibitem{Bergeret4}
R.S. Keizer, S.T.B. Goennenwein, T.M. Klapwijk, G. Miao, G. Xiao, and A. Gupta, 
Nature \textbf{439}, 825 (2006).

\bibitem{Bergeret5}
Y. V. Fominov, A. A. Golubov, and M. Y. Kupriyanov, 
JETP Lett. {\bf 77}, 510 (2003).

\bibitem{Bergeret6}
Y. Asano, Y. Tanaka, and A. A. Golubov,  
Phys. Rev. Lett. {\bf 98}, 107002 (2007). 



\bibitem{Bulut}
N. Bulut, D. J. Scalapino, and S. R. White,
Phys. Rev. B {\bf 47}, 14599 (1993).

\bibitem{Bickers}
N. E. Bickers, D. J. Scalapino, and S. R. White,
Phys. Rev. Lett. {\bf 62}, 961 (1989).

\bibitem{Kondo}
H. Kondo and T. Moriya,
J. Phys. Soc. Jpn. {\bf 67}, pp.3695 (1998).

\bibitem{NaCo-kuroki}
K. Kuroki, S. Onari, Y. Tanaka, R. Arita, and T. Nojima,
Phys. Rev. B {\bf 73}, 184503 (2006).

\bibitem{NaCo-motizuki}
M. Mochizuki, Y. Yanase, and M. Ogata, 
Phys. Rev. Lett. {\bf 94}, 147005 (2005).

\bibitem{SRO-onari}
S. Onari, R. Arita, K. Kuroki, and H. Aoki, 
J. Phys. Soc. Jpn. {\bf 74}, 2579 (2005).

\bibitem{organic-kuroki}
K. Kuroki, T. Kimura, R. Arita, Y. Tanaka, and Y. Matsuda, 
Phys. Rev. B {\bf 65}, 100516(R) (2002).

\bibitem{organic-kontani}
H. Kino and H. Kontani, 
J. Phys. Soc. Jpn. {\bf 67}, 3691 (1998).

\bibitem{High-Tc}
P. Monthoux and D. J. Scalapino, 
Phys. Rev. Lett. {\bf 72}, 1874 (1994).

\bibitem{High-Tc1}
T. Dahm and L. Tewordt, 
Phys. Rev. B {\bf 52}, 1297 (1995).

\bibitem{High-Tc2}
S. Koikegami, S. Fujimoto, and K. Yamada, 
J. Phys. Soc. Jpn. {\bf 66}, 1438 (1997).

\bibitem{High-Tc3}
T. Takimoto and T. Moriya, 
J. Phys. Soc. Jpn. {\bf 66}, 2459 (1997).

\bibitem{disconnect-kuroki}
K. Kuroki and R. Arita, 
Phys. Rev. B {\bf 64}, 024501 (2001).

\bibitem{disconnect-kuroki1}
K. Kuroki, Y. Tanaka, and R. Arita, 
Phys. Rev. Lett. {\bf 93}, 077001 (2004).


\bibitem{disconnect-onari}
S. Onari, R. Arita, K. Kuroki, H. Aoki, 
Phys. Rev. B {\bf 68}, 024525 (2003).


\bibitem{Fuseya1}
Y. Fuseya and Y. Suzumura, 
J. Phys. Soc. Jpn. {\bf 74}, 1263 (2005).

\bibitem{Fuseya11}
J. C. Nickel, R. Duprat, C. Bourbonnais, and N. Dupuis, 
Phys. Rev. Lett. {\bf 95}, 247001 (2005).

\bibitem{Fuseya12}
Y. Tanaka and K. Kuroki, 
Phys. Rev. B {\bf 70}, 060502(R) (2004).

\bibitem{Fuseya13}
K. Kuroki and Y. Tanaka, 
J. Phys. Soc. Jpn. {\bf 74}, 1694 (2005).

\bibitem{Fuseya14}
H. Aizawa, K. Kuroki, T. Yokoyama, and Y. Tanaka, 
arXiv:0811.4213. 

\bibitem{TK}
C. C. Tsuei and J. R. Kirtley, 
Rev. Mod. Phys. {\bf 72}, 969 (2000).


\bibitem{TK1}
Y. Tanaka and S. Kashiwaya, 
Phys. Rev. Lett. {\bf 74}, 3451 (1995).

\bibitem{TK2}
S. Kashiwaya and Y. Tanaka, 
Rep. Prog. Phys. {\bf 63}, 1641 (2000).

\bibitem{Linder}
J. Linder, T. Yokoyama, Y. Tanaka, Y. Asano, and A. Sudbo, 
Phys. Rev. B {\bf 77}, 174505 (2008).

\bibitem{Linder1}
J. Linder, T. Yokoyama, and A. Sudbo, 
Phys. Rev. B {\bf 77}, 174507 (2008).

\bibitem{Linder2}
Ya. V. Fominov, 
JETP Lett. {\bf 86}, 732 (2007).

\end{thebibliography}
\end{document}